\documentclass[reprint,aip,jcp,amsmath,citeautoscript]{revtex4-1}

\usepackage{graphicx}
\usepackage{txfonts}
\usepackage{bm}
\usepackage{color}
\usepackage{braket}
\usepackage{soul}
\usepackage{verbatim}
\usepackage{multirow}
\usepackage{array}
\definecolor{myblue}{rgb}{0,0,1}
\usepackage[breaklinks=true,colorlinks=true,linkcolor=myblue,urlcolor=myblue,citecolor=myblue]{hyperref}

\begin{document}


\title{Mixed Quantum--Classical Dynamics Yields Anharmonic Rabi Oscillations}

\author{Ming-Hsiu Hsieh}
\author{Roel Tempelaar}
\email{roel.tempelaar@northwestern.edu}

\affiliation{Department of Chemistry, Northwestern University, 2145 Sheridan Road, Evanston, Illinois 60208, USA}

\begin{abstract}
We apply a mixed quantum--classical (MQC) approach to the quantum Rabi model, involving a classical optical field coupled self-consistently to a quantum two-level system. Under the rotating wave approximation, we analytically show this approach to yield persistent yet anharmonic Rabi oscillations, governed by an undamped and unforced Duffing equation. We consider the single-quantum limit, where we find such anharmonic Rabi oscillations to closely follow full-quantum results once zero-point energy is approximately enforced when initializing the optical field coordinate. Our findings provide guidance in the application of MQC dynamics to classes of problems involving small quantum numbers and far away from decoherence.
\end{abstract}

\maketitle

\section{Introduction}

Mixed quantum--classical (MQC) dynamics offers a powerful framework for the simulation and interpretation of various phenomena. Throughout the last few decades, MQC dynamics has established itself as a preferred approach for the low-cost simulation of molecular systems, wherein nuclear motion is represented by classical coordinates while a quantum treatment is reserved for the electronic states \cite{tullyMixed1998, nelsonNonadiabatic2014, subotnikUnderstanding2016, wangRecent2016, curchodAbInitio2018, crespo-oteroRecent2018}. In recent years, it has been similarly employed for the modeling of materials \cite{nieUltrafast2014, smithModeling2020, xiangRealTime2021, livelyRevealing2024, krotzMixed2024} and condensed-phase correlated phenomena \cite{petrovicTheoretical2024}.

Although MQC dynamics is commonly found to provide accurate, or at least predictive, results, its realm of applicability remains poorly understood. According to the Bohr correspondence principle \cite{bohrUber1920}, the classical approximation taken within MQC dynamics is justified so long as the relevant coordinates attain large quantum numbers. However, MQC dynamics has been shown to reach high accuracy even in cases where this criterion is grossly violated \cite{tempelaarGeneralization2017, bondarenkoOvercoming2023}. Such favorable performance may be explained based on the quasiclassical dynamics literature, where it has been shown that classical equations of motion reproduce quantum dynamics for harmonic potentials and in the short-time limit \cite{hellerFrozen1981, hermanSemiclassical1984, kayHerman2006}. Yet, counterintuitive behaviors may arise once quantum and classical equations of motion are combined within a same framework, as famously exemplified by the Schr\"odinger cat thought experiment \cite{schrodingerGegenwartige1935}. As an alternative, one may therefore resort to decoherence as a criterion for the applicability of MQC dynamics, meaning that the quantum--classical correlations should remain limited.

While most applications of MQC dynamics involve electron--nuclear coupling, light--matter interactions provide a complementary realm of applicability \cite{sakkoDynamical2014, ruggenthalerFrom2018, hoffmannCapturingVacuumFluctuations2019, hoffmannBenchmarkingSemiclassicalPerturbative2019, hoffmannEffectManyModes2020, hsiehMeanFieldTreatmentVacuum2023, hsiehEhrenfestModelingCavity2023, zhangLight2023}. Notably, the canonical Rabi model, involving a two-level system (TLS) interacting with a single optical mode, was originally proposed within a MQC treatment, invoking a classical approximation for the mode \cite{rabiSpace1937}. While originally investigated under high field intensities and with neglect of feedback of the TLS onto the mode, a fully self-consistent treatment has since been explored \cite{grahamTwoState1984, cives-esclopInfluence1999}, as well as behaviors in the quantum--classical crossover region \cite{twyeffort-irishDefining2022}. However, rapid advances in the areas of quantum optics \cite{reisererCavity2015} and cavity quantum electrodynamics \cite{ebbesenHybrid2016, ribeiroPolaritonChemistryControlling2018, baranovNovel2018, hertzogStrongLightMatter2019, hiraiRecentProgress2020, dunkelbergerVibrationCavityPolaritonChemistry2022, mandalTheoreticalAdvances2023, xiangMolecular2024} have generated a particular interest in phenomena where field intensities approach the single-photon limit, as governed by the quantum Rabi model. Much remains to be learned about the applicability of MQC dynamics in this limit, and the ability of the quantum Rabi model to expose general guiding principles for MQC modeling.

Here, we apply MQC dynamics to the quantum Rabi model, specifically considering a TLS in its excited state coupled self-consistently to a single resonant optical mode prepared in its ground state. When represented fully quantum-mechanically, this model is known to exhibit Rabi oscillations manifested as harmonic population fluctuations of the TLS \cite{allenOptical2012}. As we show analytically within the rotating wave approximation (RWA), upon adopting a classical representation for the optical mode, population oscillations are instead governed by an undamped and unforced Duffing equation \cite{duffingErzwungene1918}. This equation involves an anharmonic term contributing a ``mode softening effect'' to the Rabi oscillations. The solutions to this equation are found to follow the quantum results closely upon (approximately) reinforcing zero-point energy when initializing the classical coordinate. These findings inform on the application of MQC dynamics to coherent phenomena involving small quantum numbers.

\section{Full-quantum representation}

\begin{figure}
    \includegraphics{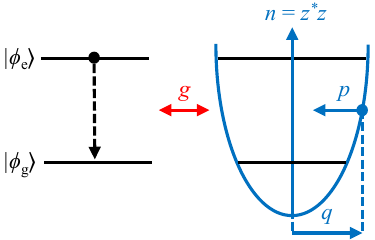}
    \caption{Schematic of the Rabi model. Shown are the ground and excited states of a two-level system (TLS), denoted $\ket{\phi_\mathrm{g}}$ and $\ket{\phi_\mathrm{e}}$, respectively. Also shown are the lowest two quantum levels of an optical mode, superimposed on a classical harmonic oscillator depiction, involving position and momentum coordinates $q$ and $p$, respectively. The classical optical occupancy, $n=z^* z$, is indicated. The coupling between the TLS and the optical mode is denoted as $g$.}
    \label{fig_scheme}
\end{figure}

Before turning to MQC dynamics, we begin by briefly revisiting the quantum Rabi model within a full-quantum representation. For simplicity, we restrict ourselves to the scenario where the transition energy of the TLS is perfectly resonant with the optical mode. A schematic of this scenario is depicted in Fig.~\ref{fig_scheme}. Setting $\hbar = 1$, and taking $\Omega$ to denote the TLS transition energy as well as the optical mode frequency, the corresponding Hamiltonian is given by
\begin{equation}
    \hat{H} = \frac{1}{2}\left(\hat{p}^2+\Omega^2\hat{q}^2\right) + \Omega \ket{\phi_\mathrm{e}}\bra{\phi_\mathrm{e}} + \mu\Omega\lambda\hat{q}\left(\ket{\phi_\mathrm{e}}\bra{\phi_\mathrm{g}} + \mathrm{H.c.}\right). \label{eq_H_quantum}
\end{equation}
Here, $\hat{q}$ and $\hat{p}$ are coordinate-like operators representing the optical mode, $\ket{\phi_\mathrm{g}}$ and $\ket{\phi_\mathrm{e}}$ refer to the ground and excited states of the TLS, $\mu$ is the associated transition dipole, $\lambda$ denotes the coupling strength between the TLS and the mode, and H.c.~is short for Hermitian conjugate.

The quantum Rabi Hamiltonian, Eq.~\ref{eq_H_quantum}, is commonly formulated by expressing the TLS in terms of Pauli spin matrices, which allows mixed states to be accounted for. Since, however, within MQC dynamics we are exclusively dealing with pure states (possibly accounting for mixed states through an ensemble average over pure states), we instead adopt a ground- and excited-state notation, which compared to the Pauli matrix formalism eliminates one variable, thereby simplifying the forthcoming analysis. Moreover, this notation enforces a connection to quantum optics and quantum electrodynamics where the TLS commonly represents the lowest two states of an atom or other quantum emitter. In both research areas, optical phenomena usually remain within the single-quantum limit, while being amenable to the RWA.

Under these conditions, Eq.~\ref{eq_H_quantum} can be simplified by first expanding the position-like operator $\hat{q}$ into the optical ladder operators,
\begin{equation}
    \hat{q} = \sqrt{\frac{1}{2\Omega}}\left(\hat{b}^\dagger + \hat{b}\right),
\end{equation}
and by introducing the effective coupling constant $g \equiv \mu\lambda\sqrt{\Omega/2}$ (see Fig.~\ref{fig_scheme}). Application of the RWA, which relies on $\Omega \gg g$, then yields a single-quantum Jaynes--Cummings (JC) Hamiltonian \cite{jaynesComparisonQuantumSemiclassical1963, shoreJaynes1993, larsonJaynes2021} of the form
\begin{equation}
    \hat{H} = \bar{\epsilon} + g\left(\ket{\phi_\mathrm{e}}\bra{\phi_\gamma} + \mathrm{H.c.}\right). \label{eq_H_JC}
\end{equation}
Here, $\ket{\phi_\mathrm{e}}$ implicitly assumes the optical mode to reside in its ground state, $\ket{0}$. The state $\ket{\phi_\gamma}$, on the other hand, denotes the first excitation of the optical mode, i.e., $\hat{b}^\dagger\ket{0}$, with the TLS in its ground state. Both $\ket{\phi_\mathrm{e}}$ and $\ket{\phi_\gamma}$ are degenerate and have an energy $\bar{\epsilon} = \frac{3}{2}\Omega$, with contributions $\Omega$ and $\frac{1}{2}\Omega$ due to the single quantum and the optical mode's zero-point energy, respectively.

Upon an initialization of the TLS in state $\ket{\phi_\mathrm{e}}$, coherent energy exchange will ensue in the form of Rabi oscillations, manifested in the transient quantum populations. For state $\ket{\phi_\mathrm{e}}$, the population is given by $P_\mathrm{e} = \vert c_\mathrm{e} \vert^2$, with the wavefunction coefficient given by
\begin{equation}
    c_\mathrm{e} = \Braket{\phi_\mathrm{e}|e^{-i\hat{H}t}|\phi_\mathrm{e}},
\end{equation}
where $t$ denotes the time elapsed after initiation. This coefficient can be evaluated straightforwardly within the eigenbasis of the JC Hamiltonian, Eq.~\ref{eq_H_JC}, with eigenvectors given by
\begin{equation}
    \ket{\psi_\pm} = \frac{1}{\sqrt{2}}\left(\ket{\phi_\mathrm{e}}\pm\ket{\phi_\gamma}\right),
\end{equation}
and with corresponding eigenenergies $\epsilon_\pm = \bar{\epsilon}\pm g$. It follows that
\begin{align}
    c_\mathrm{e} &= \frac{1}{2}\left(\Braket{\psi_+|e^{-i\epsilon_+t}|\psi_+} + \Braket{\psi_-|e^{-i\epsilon_-t}|\psi_-}\right) \nonumber \\
    &= e^{-i\bar{\epsilon} t}\cos(gt),
\end{align}
yielding
\begin{equation}
    P_\mathrm{e} = \frac{1}{2} + \frac{1}{2}\cos\left(2gt\right), \label{eq_pop_quantum}
\end{equation}
which exposes harmonic Rabi oscillations at a frequency of $2g$.

Upon a replacement of the optical mode by a classical oscillator, transient quantum populations can no longer be conveniently evaluated in the eigenbasis of the quantum Hamiltonian, and are to be evaluated in the physical basis instead. In order to contrast the resulting approach with that taken for the full-quantum representation, it is worth re-evaluating the latter within the physical basis. Accordingly, we introduce the wavefunction of the TLS, which is expanded as
\begin{equation}
    \ket{\Psi} = c_\mathrm{e}\ket{\phi_\mathrm{e}} + c_\gamma \ket{\phi_\gamma},
\end{equation}
and whose evolution is governed by the time-dependent Schr\"odinger equation, $i\ket{\dot{\Psi}
} = \hat{H}\ket{\Psi}$. This yields dynamical equations for the wavefunction coefficients given by
\begin{equation}
    \dot{c}_\mathrm{e} = -i\bar{\epsilon} c_\mathrm{e} - ig c_\gamma, \qquad
    \dot{c}_\gamma = -i\bar{\epsilon} c_\gamma - ig c_\mathrm{e}.
    \label{eq_c_dot}
\end{equation}

When $\bar{\epsilon} \gg g$ (which is equivalent to $\Omega \gg g$), the evolution of the wavefunction coefficients will be dominated by phase oscillations at a frequency $\bar{\epsilon}$, due to the diagonal contributions to Eq.~\ref{eq_c_dot}. Notably, these phase oscillations do not contribute to the population dynamics, which is instead dominated by a slower dynamical component. It proves therefore convenient to extract the slower component by dividing out the phase oscillations. Accordingly, we take
\begin{equation}
    c_\mathrm{e} = \tilde{c}_\mathrm{e} e^{-i\bar{\epsilon} t}, \qquad c_\gamma = \tilde{c}_\gamma e^{-i\bar{\epsilon} t},
    \label{eq_decompose}
\end{equation}
where the slower component is denoted with a tilde. We note that, in principle, this decomposition can be performed even when $\bar{\epsilon} \gg g$ is violated. By application of the product rule for differentiation, it follows that
\begin{equation}
    \dot{c}_\mathrm{e} = \dot{\tilde{c}}_\mathrm{e} e^{-i\bar{\epsilon} t} - i\bar{\epsilon} \tilde{c}_\mathrm{e} e^{-i\bar{\epsilon} t},\qquad \dot{c}_\gamma = \dot{\tilde{c}}_\gamma e^{-i\bar{\epsilon} t} - i\bar{\epsilon} \tilde{c}_\gamma e^{-i\bar{\epsilon} t}.
    \label{eq_c_dot_alt}
\end{equation}
Combining Eqs.~\ref{eq_c_dot}-\ref{eq_c_dot_alt} yields
\begin{equation}
    \dot{\tilde{c}}_\mathrm{e} = -ig\tilde{c}_\gamma, \qquad \dot{\tilde{c}}_\gamma = -ig\tilde{c}_\mathrm{e},
\end{equation}
upon which we find
\begin{equation}
    \ddot{\tilde{c}}_\mathrm{e} = -g^2\tilde{c}_\mathrm{e}.
    \label{eq_c_ddot}
\end{equation}
Under the initial conditions $\tilde{c}_\mathrm{e} = 1$ and $\dot{\tilde{c}}_\mathrm{e} = 0$, this is solved by $\tilde{c}_\mathrm{e} = \cos(gt)$, from which the transient excited-state population of the TLS follows as
\begin{equation}
    P_\mathrm{e} = \tilde{c}_\mathrm{e}^2 = \frac{1}{2} + \frac{1}{2}\cos\left(2gt\right),
\end{equation}
in agreement with Eq.~\ref{eq_pop_quantum}.

\section{Mixed quantum--classical representation}

We now proceed by taking the classical approximation for the optical mode, starting from the quantum Rabi model. Accordingly, we replace the coordinate-like operators by their classical canonical coordinates in Eq.~\ref{eq_H_quantum}, i.e., $\hat{p}\rightarrow p$ and $\hat{q}\rightarrow q$. This substitution yields a MQC representation in which the optical mode is modeled by a classical harmonic oscillator while its interaction with the TLS remains self-consistent, as depicted in Fig.~\ref{fig_scheme}.\footnote{Here, we avoid the commonly used ``semiclassical'' denomination, in order to differentiate such treatment from that of a quasiclassical approach, for which similar terminology is oftentimes used.} In the following, it proves convenient to express the optical mode in terms of a complex-valued classical coordinate defined as \cite{krotzReciprocal2021, miyazakiMixed2024}
\begin{equation}
    z \equiv \sqrt{\frac{\Omega}{2}}\left(q + i \frac{p}{\Omega} \right).
\end{equation}
Accordingly, we arrive at a MQC Hamiltonian of the form
\begin{equation}
    \hat{H} = \Omega z^* z + \Omega \ket{\phi_\mathrm{e}}\bra{\phi_\mathrm{e}} + g\left(z^* + z\right)\left(\ket{\phi_\mathrm{e}}\bra{\phi_\mathrm{g}} + \mathrm{H.c.}\right).
    \label{eq_H_JC_MQC}
\end{equation}

We will now follow an approach similarly to that presented above for the full-quantum representation in the physical basis. First, the dynamical equations for the wavefunction coefficients take the form
\begin{align}
    \dot{c}_\mathrm{e} &= -i(n+1)\Omega c_\mathrm{e} - ig\left(z^* + z\right) c_\mathrm{g}, \nonumber \\
    \dot{c}_\mathrm{g} &= -in\Omega c_\mathrm{g} - ig\left(z^* + z\right) c_\mathrm{e},
    \label{eq_c_dot_MQC}
\end{align}
where we introduced the (classical) optical occupancy $n\equiv z^* z$, which is representative of the optical field intensity.

As before, we divide out the phase oscillations from the wavefunction coefficients. However, according to Eq.~\ref{eq_c_dot_MQC}, the frequency of these phase oscillations is now dependent on $n$, which itself evolves in time. Accordingly, we take
\begin{equation}
    c_\mathrm{e} = \tilde{c}_\mathrm{e}e^{-i\Omega\int n\; \mathrm{d}t}e^{-i\Omega t}, \qquad c_\mathrm{g} = \tilde{c}_\mathrm{g}e^{-i\Omega\int n\; \mathrm{d}t},
    \label{eq_decompose_MQC}
\end{equation}
where the integrals in the exponent are taken over the interval from initiation to the time at which the wavefunction coefficients are being evaluated. The equivalent of Eq.~\ref{eq_c_dot_alt} now becomes
\begin{align}
    \dot{c}_\mathrm{e} &= \dot{\tilde{c}}_\mathrm{e}e^{-i\Omega\int n\; \mathrm{d}t}e^{-i\Omega t} - i(n+1)\Omega c_\mathrm{e}, \nonumber \\
    \dot{c}_\mathrm{g} &= \dot{\tilde{c}}_\mathrm{g}e^{-i\Omega\int n\; \mathrm{d}t} - in\Omega c_\mathrm{g}.
    \label{eq_c_dot_alt_MQC}
\end{align}
Combining Eqs.~\ref{eq_c_dot_MQC}-\ref{eq_c_dot_alt_MQC} yields
\begin{align}
    \dot{\tilde{c}}_\mathrm{e}e^{-i\Omega t} &= - ig\left(z^* + z\right) \tilde{c}_\mathrm{g}, \nonumber \\
    \dot{\tilde{c}}_\mathrm{g} &= - ig\left(z^* + z\right) \tilde{c}_\mathrm{e} e^{-i\Omega t}.
    \label{eq_almost_there}
\end{align}

Our goal is to reformulate Eq.~\ref{eq_almost_there} into a single, closed-form equation for $\tilde{c}_\mathrm{e}$, similarly to Eq.~\ref{eq_c_ddot} for the full-quantum case. To this end, we need to eliminate $z$ and $\tilde{c}_\mathrm{g}$. We first recognize that the dynamics of $z$ is governed by \cite{miyazakiMixed2024}
\begin{equation}
    \dot{z} = -i\frac{\partial H}{\partial z^*}.
\end{equation}
Adopting the Ehrenfest theorem \cite{ehrenfestBemerkung1927}, the classical Hamiltonian follows from the quantum Hamiltonian as
\begin{align}
    H &= \braket{\Psi|\hat{H}|\Psi} \nonumber \\
    &= \Omega z^* z + \Omega \tilde{c}_\mathrm{e}^*\tilde{c}_\mathrm{e} + g(z^*+z)\left(c_\mathrm{e}^* c_\mathrm{g} + \mathrm{H.c.}\right).
    \label{eq_Ehrenfest}
\end{align}
This yields
\begin{equation}
    \dot{z} = -i\Omega z - ig\left(c_\mathrm{e}^* c_\mathrm{g} + \mathrm{H.c.}\right).
    \label{eq_z_dot}
\end{equation}
Similarly as for the wavefunction coefficients, it proves convenient to divide phase oscillations out of $z$, according to
\begin{equation}
    z = \tilde{z}e^{-i\Omega t}.
\end{equation}
Substitution into Eq.~\ref{eq_almost_there} yields
\begin{equation}
    \dot{\tilde{c}}_\mathrm{e} = - ig\tilde{z} \tilde{c}_\mathrm{g}, \qquad \dot{\tilde{c}}_\mathrm{g} = - ig\tilde{z}^* \tilde{c}_\mathrm{e},
\end{equation}
where we adopted the RWA as we did for the full-quantum representation. From this, it follows that
\begin{align}
    \ddot{\tilde{c}}_\mathrm{e} &= - ig(\tilde{z}\dot{\tilde{c}}_\mathrm{g} + \dot{\tilde{z}} \tilde{c}_\mathrm{g}) \nonumber \\
    &= - g^2 n \tilde{c}_\mathrm{e}
    - ig\dot{\tilde{z}} \tilde{c}_\mathrm{g},
    \label{eq_almost_almost_there}
\end{align}
where we used that the optical occupancy can be expressed as $n=\tilde{z}^*\tilde{z}$.

Evaluation of $\dot{\tilde{z}}$ proceeds in the same way as for the wavefunction coefficients. Accordingly, we first recognize that
\begin{equation}
    \dot{z} = \dot{\tilde{z}}e^{-i\Omega t} - i\Omega z,
\end{equation}
which combined with Eqs.~\ref{eq_c_dot_alt_MQC} and \ref{eq_z_dot} yields
\begin{equation}
    \dot{\tilde{z}} = - ig\tilde{c}_\mathrm{g}^*\tilde{c}_\mathrm{e}.
    \label{eq_z_tilde_dot}
\end{equation}
Substitution into Eq.~\ref{eq_almost_almost_there} results in
\begin{equation}
    \ddot{\tilde{c}}_\mathrm{e} = -g^2\left(1 + n - \tilde{c}_\mathrm{e}^*\tilde{c}_\mathrm{e}\right) \tilde{c}_\mathrm{e},
    \label{eq_c_dot_cod_MQC_preliminary}
\end{equation}
where we employed the normalization condition $\tilde{c}_\mathrm{e}^*\tilde{c}_\mathrm{e} + \tilde{c}_\mathrm{g}^*\tilde{c}_\mathrm{g} = 1$ in order to eliminate $\tilde{c}_\mathrm{g}^*\tilde{c}_\mathrm{g}$.

So long as $\Omega \gg g$ is satisfied, the total combined energy of the TLS and the optical field is well approximated as $E = N\Omega$, with $N = \left(n + \tilde{c}_\mathrm{e}^* \tilde{c}_\mathrm{e}\right)$ representing the sum of the occupancy of the TLS excited state and that of the optical field. Since $E$ is a conserved quantity, so is $N$. If we initialize the TLS in its excited state, i.e., $\tilde{c}_\mathrm{e} = 1$, and denote the initial optical occupancy as $n_0$, we thus have that $n + \tilde{c}_\mathrm{e}^2 = n_0 + 1$. Moreover, initializing $\tilde{c}_\mathrm{e}$ as real-valued ensures this coefficient will remain real-valued throughout the dynamics. Consequently, Eq.~\ref{eq_c_dot_cod_MQC_preliminary} simplifies to
\begin{equation}
    \ddot{\tilde{c}}_\mathrm{e} =-\left(2 + n_0\right)g^2 \tilde{c}_\mathrm{e} + 2g^2\tilde{c}_\mathrm{e}^3.
    \label{eq_c_ddot_MQC}
\end{equation}

Eq.~\ref{eq_c_ddot_MQC} takes the form of an undamped and unforced Duffing equation \cite{duffingErzwungene1918}, with a harmonic term $-(2+n_0)g^2$ and an anharmonic term $2g^2$. The latter marks a difference in behavior of MQC dynamics compared to the full-quantum representation, the latter being characterized by purely-harmonic dynamics (cf.~Eq.~\ref{eq_c_ddot}). The anharmonicity arises from the self-consistent interaction between the TLS and the optical field under the classical approximation taken for the latter. Specifically, the classical approximation introduces a parametric dependence of the quantum Hamiltonian on the classical coordinate, while the classical equations of motion receive a feedback contribution proportional to $\tilde{c}_\mathrm{e}^2$, ultimately yielding the $\tilde{c}_\mathrm{e}^3$ term in Eq.~\ref{eq_c_ddot_MQC}. (We note that the anharmonicity disappears when feedback of the TLS onto the classical coordinates is neglected.)

The Duffing equation is ubiquitous in many branches of physics and engineering, and has long been under scrutiny \cite{kovacicDuffing2011, cveticaninNinety2013}. The particular form given by Eq.~\ref{eq_c_ddot_MQC} is solved in terms of Jacobi elliptic functions, with specific solutions dependent on the initial conditions taken. Insightful simplified forms of these solutions have remained out of reach, but what is known is that the solutions are periodic \cite{abramowitzHandbook1968}. As such, we find persistent oscillations to be retained under MQC dynamics, showing a dependency on the initial optical occupancy, $n_0$. Within this oscillatory motion, the anharmonic term in the Duffing equation yields a reduction of the ``restoring force'' with increasing amplitude of $\tilde{c}_\mathrm{e}$, corresponding to an effective decrease in mode stiffness. For this reason, the anharmonic term is said to exert a ``mode softening'' effect on the Rabi oscillations.

There is a trivial limit for which a simple analytical solution to Eq.~\ref{eq_c_ddot_MQC} is readily available, corresponding to $n_0 = 0$. In this limit, Eq.~\ref{eq_c_ddot_MQC} reduces to $\ddot{\tilde{c}}_\mathrm{e} \approx 0$. If we reinforce $\dot{\tilde{c}}_\mathrm{e} = 0$ as an initial condition, as before, we thus find a complete absence of dynamics in this limit. This result is consistent with the fundamental principle that spontaneous emission requires optical vacuum fluctuations to be incorporated \cite{liMixedQuantumclassicalElectrodynamics2018}. These are not automatically accounted for by classical coordinates, and vanish rigorously for $n_0 = 0$.

Incorporation of optical vacuum fluctuations thus requires an initialization satisfying $n_0 > 0$. There are various possibilities for which this criterion can be satisfied. For now, however, we chose to remain agnostic to these possibilities, and instead proceed to systematically and numerically assess the effect of classical coordinate initialization on the Rabi oscillations governed by Eq.~\ref{eq_c_ddot_MQC}. Notably, through the dependence of this equation on the initial optical occupancy, $n_0 = z_0^* z_0$, these oscillations depend on the norm of the initial coordinate $z_0$, but not on its phase. This is a direct consequence of the condition $\Omega \gg g$, implying that phase oscillations are infinitely rapid compared to the coherent energy exchange between the TLS and the optical mode.

\begin{figure}
    \includegraphics{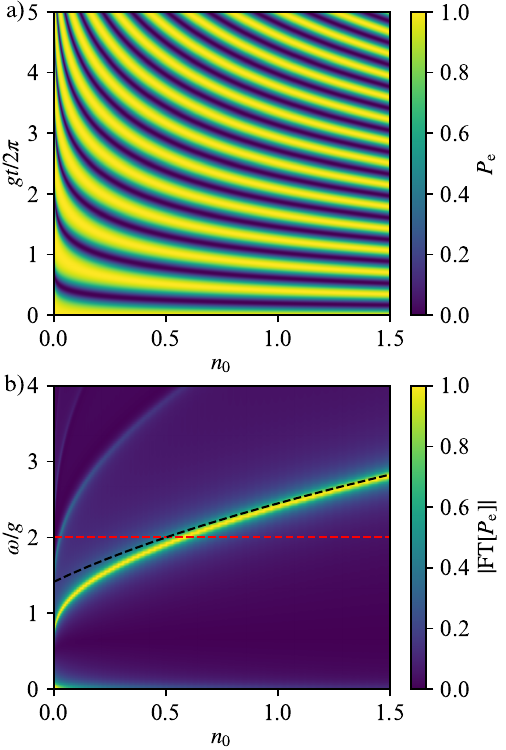}
    \caption{(a) Transient excited-state population of the TLS, $P_\mathrm{e} = \tilde{c}_\mathrm{e}^2$, obtained from the solutions of the undamped and unforced Duffing equation, Eq.~\ref{eq_c_ddot_MQC}. Shown are results as a function of the initial optical occupancy, $n_0$. (b) Fourier transform of $P_\mathrm{e}$. Asymptotic frequency of the dominant harmonic given by Eq.~\ref{eq_asymptote} (black dash) and quantum Rabi oscillation frequency at $\omega = 2g$ (red dash) are indicated.}
    \label{fig_scan}
\end{figure}

We will proceed to assess the dynamics produced by the Duffing equation in terms of the excited-state population of the TLS, $P_\mathrm{e} = \tilde{c}_\mathrm{e}^2$. Other observables of interest include the TLS ground-state population, which due to normalization follows from the excited-state population as $\tilde{c}_\mathrm{g}^2 = 1 - \tilde{c}_\mathrm{e}^2$, and the optical occupancy, which through energy conservation follows as $n = E/\Omega - \tilde{c}_\mathrm{e}^2$. As such, these observables and their behaviors can trivially be inferred from the excited-state population data presented.

Shown in Fig.~\ref{fig_scan} (a) is $P_\mathrm{e}$ obtained from the solutions to Eq.~\ref{eq_c_ddot_MQC}, as a function of time and of $n_0$. As anticipated, clear periodic oscillations are observed. Evidently, the oscillation frequency is proportional to $n_0$, suggesting that the lack of dynamics in the limit $n_0\rightarrow 0$ can be understood as an oscillation with infinite periodicity. The frequency trend is further elucidated in Fig.~\ref{fig_scan} (b), showing the Fourier transform of $P_\mathrm{e}$ as a function of frequency, $\omega$, and initial optical occupancy, $n_0$. For each $n_0$, the Fourier transform was performed over a duration of $gt = 200$, after multiplication of $P_\mathrm{e}$ by an exponential with an $1/e$ time of $gt = 20$ to avoid clipping and after subtracting a 0.5 baseline, upon which the signal was normalized to the maximum peak. A dominant oscillatory component is observable alongside several minor components that vanish as $n_0$ increases. This is suggestive of harmonic oscillatory motion being recovered for $n_0\gg 0$. This limit turns out to be amenable to an approximate analytical treatment, providing deeper insights into Rabi oscillations emerging under MQC dynamics.

Accordingly, we first recognize that for $n_0\gg 0$ the anharmonic term will provide only a minor contribution to Eq.~\ref{eq_c_ddot_MQC}. As a result, we may expect the solution to this equation to be well-represented by the ansatz $\tilde{c}_\mathrm{e} = \cos(\tilde{\omega} t)$, where $\tilde{\omega}$ is the dominant harmonic that needs determining. Substituting this ansatz into Eq.~\ref{eq_c_ddot_MQC} yields
\begin{align}
    -\tilde{\omega}^2 \cos(\tilde{\omega} t) &= -(2 + n_0)g^2 \cos(\tilde{\omega} t) + \frac{3}{2} g^2 \cos(\tilde{\omega} t) \nonumber \\
    & \qquad + \frac{1}{2}g^2 \cos(3 \tilde{\omega} t),
    \label{eq_c_ddot_substituted}
\end{align}
where we used that
\begin{equation}
    \cos^3(\tilde{\omega}t) = \frac{3}{4} \cos(\tilde{\omega} t) +  \frac{1}{4}\cos(3\tilde{\omega} t).
\end{equation}

The $\cos(3 \tilde{\omega} t)$ term appearing on the right-hand side of Eq.~\ref{eq_c_ddot_substituted} deserves a brief discussion. First, its prefactor is $\times 3$ smaller than that of the preceding term, which in turn is already a correction to the first term involving $n_0$. Hence, although our ansatz can be refined in order to include the $\cos(3 \tilde{\omega} t)$ contribution, we expect resulting corrections to the dominant oscillation frequency to be minor. Instead, we proceed with Eq.~\ref{eq_c_ddot_substituted} and balance only the $\cos(\tilde{\omega} t)$ terms. Doing so, and dividing by $\cos(\tilde{\omega} t)$, yields
\begin{equation}
    \tilde{\omega}^2 \approx \left(n_0 + \frac{1}{2}\right)g^2.
\end{equation}
The asymptotic Rabi oscillation frequency then follows as
\begin{equation}
    \omega = 2\tilde{\omega} \approx 2g \sqrt{n_0 + 1/2}. \label{eq_asymptote}
\end{equation}
This asymptote is indicated in Fig.~\ref{fig_scan}, and the dominant component predicted by Eq.~\ref{eq_c_ddot_MQC} is seen to indeed approach this frequency with increasing $n_0$.

The asymptotic frequency given by Eq.~\ref{eq_asymptote} forms an interesting contrast with the known behavior arising under a full-quantum treatment of the quantum Rabi model under the JC Hamiltonian, for which we have $\omega = 2g\sqrt{n_0 + 1}$ (with $n_0$ representing the quantum occupancy of the optical field). Interestingly, the MQC Rabi model with neglect of feedback of the TLS on the classical optical field instead yields $\omega = 2g\sqrt{n_0}$ \cite{allenOptical2012}. As such, our self-consistent MQC approach presents itself as an interesting ``intermediate'' between those limiting cases.

Another curious aspect of Eq.~\ref{eq_asymptote} is that it predicts $n_0 = 1/2$ to be the value for which MQC dynamics reproduces quantum Rabi oscillations at $\omega = 2g$. For this occupancy, the classical field assumes an energy corresponding to that of vacuum fluctuations. This suggests that for MQC dynamics to reproduce the full-quantum result, the optical field needs to be initialized roughly at the zero-point energy. It should be noted, however, that Eq.~\ref{eq_asymptote} was derived assuming $n_0 \gg 0$, and loses validity at small $n_0$ values. Indeed, as seen in Fig.~\ref{fig_scan} (b), the dominant component predicted by Eq.~\ref{eq_c_ddot_MQC} begins to undershoot the asymptotic frequency with decreasing occupancy, in order to satisfy $\omega \rightarrow 0$ for $n_0 \rightarrow 0$. Consequently, MQC dynamics is found to reproduce the quantum Rabi oscillation frequency (also indicated in Fig.~\ref{fig_scan} (b)) most closely at an occupancy exceeding the zero-point value, that is, at approximately $n_0 = 0.59$.

\begin{figure}
  \includegraphics{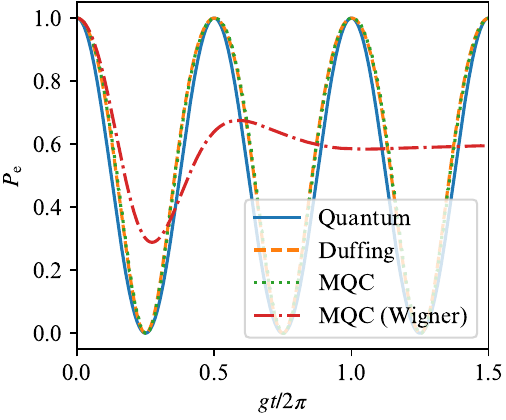}
  \caption{Excited-state population of the TLS, $P_\mathrm{e}=\tilde{c}_\mathrm{e}^2$, obtained from the solution to the undamped and unforced Duffing equation, Eq.~\ref{eq_c_ddot_MQC}, for $n_0 = 0.59$ (orange dash), compared with the full-quantum result (blue solid). Also shown are results from mixed quantum--classical (MQC) dynamics under focused sampling (dotted) and under Wigner sampling (red dash-dotted), obtained using $\Omega = 50g$.}
  \label{fig_compare}
\end{figure}

To assess how closely MQC dynamics reproduces the full-quantum result at $n_0 = 0.59$, we present in Fig.~\ref{fig_compare} a comparison of $P_\mathrm{e} = \tilde{c}_\mathrm{e}^2$ obtained through Eqs.~\ref{eq_pop_quantum} and \ref{eq_c_ddot_MQC}, as a function of time. As can be seen, both transients are in broad agreement, with a matching dominant oscillation component. Notably, we observe a population bias of the Rabi oscillation towards the excited state of the TLS, which is a consequence of the anharmonic term in Eq.~\ref{eq_c_ddot_MQC} reducing the restoring force with increasing amplitude of $\tilde{c}_\mathrm{e}$.

Fig.~\ref{fig_compare} also includes MQC dynamics results obtained by directly solving Eq.~\ref{eq_H_JC_MQC} for $n_0 = 0.59$. We note that these results do not adopt the RWA. By setting $\Omega = 50 g$, however, we ensure that inaccuracies introduced upon an adoption of the RWA, as done in the Duffing equation and the full-quantum representation, are negligible, while also ensuring that the initial phase of the optical field does not affect the dynamics. The latter justifies an initialization of the classical coordinates at $p=0$ and $q>0$, and negates the need for trajectory sampling. The results coming out of this approach are in close agreement with those produced by Eq.~\ref{eq_c_ddot_MQC} substantiating the validity of our analysis.

Our findings suggest that for the model at hand, MQC dynamics are best performed with an initialization of the classical coordinates with a well-defined energy lying slightly above the zero-point energy, by setting $n_0 = 0.59$. This contrasts with the approach commonly taken in MQC dynamics wherein initial classical coordinates are stochastically sampled from a Wigner quasi-probability distribution to represent the vacuum wavefunction of the field \cite{wignerOn1932}, giving rise to a distribution of initial energies the average of which matches the zero-point energy. To further investigate this contrast, we have also included in Fig.~\ref{fig_compare} MQC results for which such Wigner sampling is conducted. Accordingly, $p$ and $q$ are stochastically drawn from the normal distribution
\begin{equation}
    \Gamma(p, q) = \frac{1}{\pi}\mathrm{exp}\left( -\frac{p^2}{\Omega} - \Omega q^2\right),
\end{equation}
while an average over 100\;000 trajectories is taken, which proved necessary to ensure convergence.

As is obvious from Fig.~\ref{fig_compare}, such Wigner sampling performs poorly in reproducing the Rabi oscillations, yielding rapidly-decaying population dynamics instead of persistent oscillations. From Fig.~\ref{fig_scan}, the rapid decay is attributed to Wigner sampling effectively taking an average over oscillatory dynamics with different $n_0$ values for which the distribution of frequencies is broad, leading the resulting oscillations to destructively interfere. In this process, $P_\mathrm{e}$ equilibrates to a value exceeding 0.5 in Fig.~\ref{fig_compare}, due to the aforementioned population bias towards the excited state of the TLS induced by the anharmonic term in the Duffing equation, Eq.~\ref{eq_c_ddot_MQC}.

The preference of enforcing $n_0$ to (roughly) match the zero-point energy, instead of a Wigner sampling of trajectories, resonates with previous studies. At times referred to as ``focused sampling'', such reinforcement of physical properties at the single-trajectory level \cite{stockFlow1999} is commonly envisioned as a means to accelerate convergence with respect to the number of trajectories \cite{bonellaSemiclassicalImplementationMapping2003, bonellaLinearizedPath2005, kimQuantumClassical2008} and has found applications in conjunction with various quasiclassical approaches \cite{mandalQuasiDiabatic2018, runesonSpinMapping2019, ananthPathIntegrals2022}. Importantly, while focused sampling enables convergence at the single-trajectory level for the Rabi model studied here, its most essential effect is the remedying of unphysical behaviors arising under Wigner sampling. It should be noted that unphysical behaviors under Wigner sampling have been observed in previous studies \cite{tempelaarGeneralization2017, bondarenkoOvercoming2023, hsiehEhrenfestModelingCavity2023}. Lastly, we should emphasize that focused sampling and Wigner sampling differ exclusively by the initialization of the classical coordinates, leaving the equations of motion identical.

\section{Discussion}

In summary, we have shown that Rabi oscillations can to a good approximation be reproduced by MQC dynamics, even in the single-quantum limit. Upon adopting the RWA, the resulting population dynamics is governed by an undamped and unforced Duffing equation, Eq.~\ref{eq_c_ddot_MQC}, involving an anharmonic term, the solutions of which consist of persistent oscillatory functions with frequencies dependent on the initial occupancy of the optical mode. Rough agreement with the quantum result is found for the case of a TLS initiated in its excited state when approximately enforcing zero-point energy for the initial optical occupancy. We note that similar agreement is borne out upon raising this occupancy by a quantum while initializing the TLS in its ground state (see the Supplementary Material). These results underscore the advantage presented by so-called focused sampling techniques \cite{stockFlow1999} in MQC dynamics over commonly-applied Wigner sampling. However, which approach is preferred may be highly susceptible to the problem at hand; something that is worthy of further inquiry.

The dynamics of the quantum Rabi model has previously been evaluated based on a self-consistent MQC approach such as presented here, but within a Pauli matrix representation \cite{cives-esclopInfluence1999}. This analysis resulted in 5 coupled differential equations, upon which solutions could be derived in the form of Jacobi elliptic functions. As such, these solutions show consistency with our analysis which instead reduces to a single differential equation. We note that the corresponding dynamics was previously assessed exclusively for larger quantum numbers \cite{cives-esclopInfluence1999}, for which the Jacobi elliptic functions become harmonic. Indeed, within the Duffing equation, we find the anharmonicity to lose significance with increasing quantum numbers, while at the same time the exact value of the optical occupancy adopted to represent zero-point energy becomes less critical. We note that the limit of large quantum numbers could be representative of laser-driven systems, and marks a stark contrast with the quantum-optical limit explored in our work.

Whereas we restricted ourselves to the scenario of a TLS interacting with a single optical mode, previous studies have considered the performance of MQC dynamics under Wigner sampling for the scenario where the number of optical modes is large \cite{hoffmannCapturingVacuumFluctuations2019, hoffmannBenchmarkingSemiclassicalPerturbative2019, SallerBenchmarkingQuasiclassicalMapping2021, hsiehMeanFieldTreatmentVacuum2023, hsiehEhrenfestModelingCavity2023}. Under application of the Ehrenfest theorem, Eq.~\ref{eq_Ehrenfest}, MQC dynamics was shown to yield quantitative inaccuracies \cite{hoffmannCapturingVacuumFluctuations2019, hoffmannBenchmarkingSemiclassicalPerturbative2019, SallerBenchmarkingQuasiclassicalMapping2021}, and at times qualitative failures \cite{hsiehMeanFieldTreatmentVacuum2023, hsiehEhrenfestModelingCavity2023}. These shortcomings are remedied by a recently-proposed scheme wherein optical vacuum fluctuations are selectively decoupled from the atomic ground state \cite{hsiehMeanFieldTreatmentVacuum2023, hsiehEhrenfestModelingCavity2023}. It will thus be of interest to revisit such a ``decoupled'' Ehrenfest approach and incorporate the focused sampling scheme adopted in the present work, in order to deliver a MQC dynamics method providing optimal accuracy for both small and large mode numbers.

Within the multi-modal case, nondegeneracies inevitably arise between the quantum transition energies and the classical mode frequencies. In the present work, we restricted our analysis to resonance conditions, as obtaining analytical expressions beyond resonance is nontrivial. However, in the Supplementary Material we compare transient excited-state populations produced numerically through MQC dynamics with full-quantum results, and find good agreement to be retained away from resonance.

Lastly, it is worth noting that the Rabi model bears similarities with other canonical scenarios, including the spin--boson model. Among other things, it will be of particular interest to extend our assessment of the applicability of MQC dynamics to systems involving couplings with increasing degrees of anharmonicities and non-linearity, following successful examples within the realm of quasi-classical modeling \cite{wuLinearNonlinearResponse2001, kryvohuzQuantumClassicalCorrespondenceResponse2005}. Such efforts will guide the application of MQC dynamics to an increasing range of phenomena.

\section*{Supplementary material}

Numerical evaluation of the off-resonant quantum Rabi model and results for a ground-state initialization of the TLS.

\section*{Acknowledgement}

The authors thank Alex Krotz, Ken Miyazaki, and Antonio Garz\'on-Ram\'irez for helpful discussions. This material is based upon work supported by the National Science Foundation under Grant No.~2145433. Research reported in this publication was supported, in part, by the International Institute for Nanotechnology at Northwestern University. M.-H.H.~gratefully acknowledges support from the Ryan Fellowship and the International Institute for Nanotechnology at Northwestern University.

\section*{Data availability}

The data that support the findings of this study are available from the corresponding author upon reasonable request. 

\bibliography{bib}

\end{document}


\title{Supplementary Material for\\Mixed Quantum--Classical Dynamics Yields Anharmonic Rabi Oscillations}

\author{Ming-Hsiu Hsieh}
\author{Roel Tempelaar}
\email{roel.tempelaar@northwestern.edu}

\affiliation{Department of Chemistry, Northwestern University, 2145 Sheridan Road, Evanston, Illinois 60208, USA}

\maketitle


\section{Off-resonant quantum Rabi model}

\begin{figure}[!h]
\includegraphics{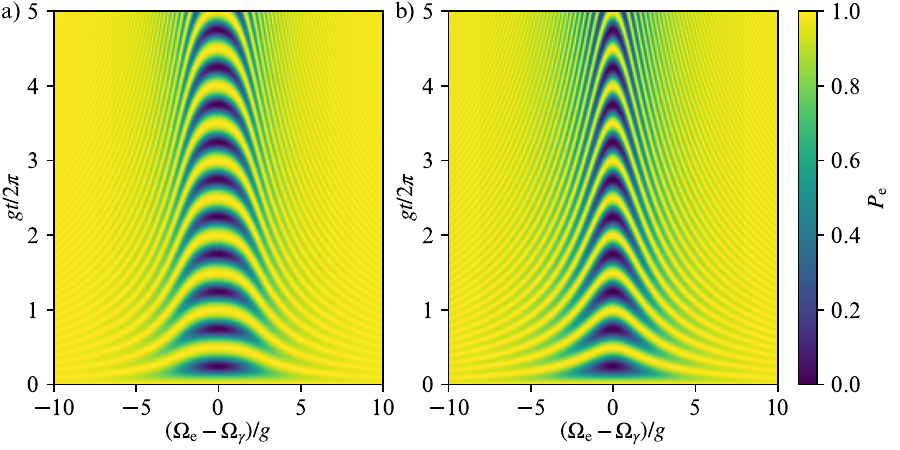}
  \caption{Transient excited-state population of the TLS, $P_\mathrm{e}=\tilde{c}_\mathrm{e}^2$, for the off-resonant quantum Rabi model. Shown are results from (a) MQC dynamics against (b) full-quantum results.}
  \label{fig_compare2}
\end{figure}

Whereas in the main text we exclusively consider the quantum Rabi model under complete resonance between the optical quantum and the transition energy of the two-level system (TLS), here we consider the case where this degeneracy is broken. Accordingly, we introduce $\Omega_\mathrm{e}$ and $\Omega_\gamma$ in order to denote the TLS transition energy and optical quantum, respectively, and vary $\Omega_\mathrm{e}$ while keeping $\Omega_\gamma$ fixed (keeping all parameters as in the main text). We note that an analytical evaluation of the resultant dynamical equations, analogously to that presented in the main text, is highly nontrivial in this generalized case. Instead, we proceed to present in Fig.~\ref{fig_compare2} a comparison of the transient excited state population of the TLS, $P_\mathrm{e}$, calculated using MQC dynamics and the full-quantum result. Here, MQC dynamics is seen to capture the trend observed while scanning across the resonant case, reaching good overall agreement with the full-quantum result.

\newpage

\section{Ground-state initialization of the two-level system}

\begin{figure}[!h]
\includegraphics{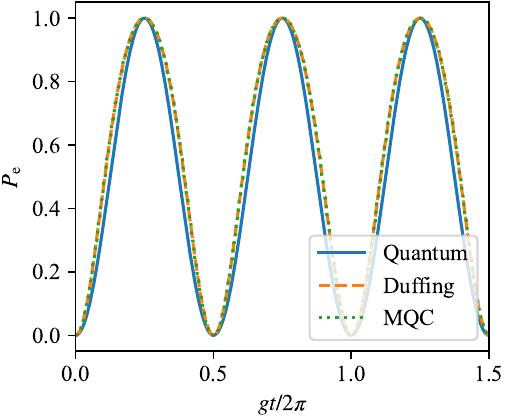}
  \caption{Same as Fig.~3 in the main text, but with a ground-state initialization of the TLS and with $n_0 = 1.59$. MQC dynamics results under Wigner sampling perform similarly poorly as in Fig.~3, and are not included here.}
  \label{fig_compare}
\end{figure}

In the main text, we found MQC dynamics to agree well with the quantum result when initializing the TLS in its excited state, provided that zero-point energy is approximately enforced for the optical field by setting $n_0=0.59$. Here, we show that a similar level of agreement is reached when initializing the TLS in its ground state, while raising the optical occupancy by one quantum, i.e., by setting $n_0 = 1.59$. In this case, Eq.~29 still holds, but the final Duffing equation (cf.~Eq.~30) now takes the form
\begin{equation}
    \ddot{\tilde{c}}_\mathrm{e} =-\left(1 + n_0\right)g^2 \tilde{c}_\mathrm{e} + 2g^2\tilde{c}_\mathrm{e}^3.
    \label{eq_Duffing_new}
\end{equation}
Shown in Fig.~\ref{fig_compare} is the equivalent of Fig.~3 in the main text, but with a ground-state initialization of the TLS and with $n_0=1.59$, confirming that both MQC dynamics and Eq.~\ref{eq_Duffing_new} agree well with the quantum result.

